\documentclass[pre,aps,a4paper,showpacs, showkeys, preprint]{revtex4}
\usepackage{amsfonts}
\usepackage{amsmath}
\usepackage{amssymb}
\usepackage{bm}

\input{tcilatex}
\begin{document}

\title{Identification of an average temperature and
 a  dynamical pressure in
a multitemperature mixture of fluids}
\author{Henri Gouin}
\email{henri.gouin@univ-cezanne.fr}
\affiliation{University of Aix-Marseille \& C.N.R.S. U.M.R.\,6181, Box 322 \\
Av. Escadrille Normandie-Niemen,   13397 Marseille Cedex 20,
France.}
\author{Tommaso Ruggeri}
\email{ruggeri@ciram.unibo.it}
\homepage[Home page: ]{http://www.ciram.unibo.it/ruggeri}
\affiliation{Department of Mathematics and Research Center of Applied Mathematics -
(C.I.R.A.M.) - \\
University of Bologna, Via Saragozza 8, 40123 Bologna, Italy.}
\date{\today}

\begin{abstract}
We present a classical approach of a mixture of compressible fluids
when each constituent has its own temperature. The introduction of
an   \emph{average temperature} together with the entropy principle
dictates  the classical Fick law for diffusion and also\emph{ novel}
constitutive equations associated with the difference of
temperatures between the components. The constitutive equations fit
with results recently obtained through   \emph{Maxwellian iteration}
procedure in extended thermodynamics theory of multitemperature
mixtures. The differences of
temperatures between the constituents imply the existence of a new \emph{%
dynamical pressure} even if the fluids have a zero bulk viscosity.
The nonequilibrium dynamical pressure can be measured and   may be
convenient in several physical situations   as for example in
cosmological circumstances where - as many authors assert - a
dynamical pressure played a major role in the evolution of the early
universe.
\end{abstract}

\keywords{Multitemperature mixture of fluids; Dynamical pressure;
Diffusion; Nonequilibrium thermodynamics.
\\
Journal:  \textbf{\qquad\qquad\qquad\quad Physical Review E 78,
n$^o$ 1, 016303 (2008)}} \pacs{51.30.+i, 47.51.+a}

\maketitle





\section{Mixtures in rational thermodynamics}

In the context of rational thermodynamics, the description of a  homogeneous mixture  of $n$ constituents is based on
the postulate that each constituent  obeys    the same balance laws that  a single
fluid does \cite{RT}.  The laws express the equations of balance of masses,
momenta and energies
\begin{equation}
\left\{
\begin{array}{ll}
& \displaystyle\frac{\partial \rho _{\alpha }}{\partial t}+\mathrm{div}%
\,(\rho _{\alpha }\mathbf{v}_{\alpha })=\tau _{\alpha }, \\
&  \\
& \displaystyle\frac{\partial (\rho _{\alpha }\mathbf{v}_{\alpha })}{%
\partial t}+\mathrm{div}\,(\rho _{\alpha }\mathbf{v}_{\alpha }\otimes
\mathbf{v}_{\alpha }-\mathbf{t}_{\alpha })=\mathbf{m}_{\alpha },\qquad
\qquad (\alpha =1,2,\dots ,n) \\
&  \\
& \displaystyle\frac{\partial \left( \frac{1}{2}\rho _{\alpha }v_{\alpha
}^{2}+\rho _{\alpha }\varepsilon _{\alpha }\right) }{\partial t}+\mathrm{div}%
\,\left\{ \left( \frac{1}{2}\rho _{\alpha }v_{\alpha }^{2}+\rho _{\alpha
}\varepsilon _{\alpha }\right) \mathbf{v}_{\alpha }-\mathbf{t}_{\alpha }%
\mathbf{v}_{\alpha }+\mathbf{q}_{\alpha }\right\} =e_{\alpha }.
\end{array}
\right.  \label{RT_model}
\end{equation}
On the left-hand side,   $\rho_\alpha$ is the density, $\mathbf{v}_{\alpha}$ is the velocity, $
\varepsilon_\alpha$ is the internal energy, $\mathbf{q}_{\alpha }$ is the heat flux and $\mathbf{t}_{\alpha}$ is the stress tensor  of the constituent  $\alpha$. The stress tensor   $\mathbf{t}_{\alpha}$ can be decomposed into a pressure part $-p_{\alpha }\mathbf{I}$ and a viscous part $\bm{\sigma}_{\alpha }$ as
\begin{equation*}
\mathbf{t}_{\alpha }=-p_{\alpha }\mathbf{I}+\bm{\sigma}_{\alpha }.
\end{equation*}
We consider first
only Stokesian fluids (\emph{i.e.} $\text{tr}\left(\bm{\sigma}_{\alpha }\right) = 0$).
\newline
On the right-hand sides  $\tau_{\alpha}$, $\textbf{m}_{\alpha}$ and $e_{\alpha}$  represent the production terms related to  the
interactions between constituents. Due to the total   conservation of mass,
momentum and energy of the mixture, the sum of production terms over all
constituents must vanish,
\begin{equation*}
\sum_{\alpha
=1}^{n}\tau_{\alpha }=0, \quad \sum_{\alpha =1}^{n}\mathbf{m}_{\alpha }= \textbf{0},\quad \sum_{\alpha
=1}^{n}e_{\alpha }=0.
\end{equation*}
For the sake of simplicity, we ignore  in the following  the possibility of chemical reactions
($\tau _{\alpha }=0$).

The mixture quantities $\rho ,\mathbf{v}, \varepsilon, \mathbf{t}$ and $%
\mathbf{q}$ are defined as
\begin{equation}
\begin{array}{ll}
\displaystyle\rho =\sum_{\alpha =1}^{n}\rho _{\alpha }, & \text{total mass
density}, \\
&  \\
\displaystyle\mathbf{v}=\frac{1}{\rho }\sum_{\alpha =1}^{n}\rho _{\alpha }%
\mathbf{v}_{\alpha }, & \text{mixture velocity}, \\
&  \\
\displaystyle\varepsilon =\displaystyle\varepsilon_I+\frac{1}{2\rho }\sum_{\alpha =1}^{n}\rho _{\alpha
}u_{\alpha }^{2}, & \text{%
internal energy, } \\
&  \\
\displaystyle\mathbf{t}=-p\mathbf{I+}\bm{\sigma}_{I}-\sum_{\alpha
=1}^{n}(\rho _{\alpha }\mathbf{u}_{\alpha }\otimes \mathbf{u}_{\alpha }), &
\text{stress tensor,} \\
&  \\
\displaystyle\mathbf{q}=\mathbf{q}_{I}+\sum_{\alpha =1}^{n}\rho _{\alpha
}\left( \varepsilon _{\alpha }+\frac{p_{\alpha }}{\rho _{\alpha }}+\frac{1}{2%
}u_{\alpha }^{2}\right) \mathbf{u}_{\alpha }, & \text{flux of internal energy},
\end{array}
\label{6}
\end{equation}%
where  $\mathbf{u}_{\alpha }=\mathbf{v}_{\alpha }-\mathbf{v}$ is the
diffusion velocity of the component $\alpha$, $p=\sum_{\alpha =1}^{n}p_{\alpha }$ is the total
pressure, $\displaystyle\varepsilon_I=\frac{1}{\rho}\sum_{\alpha =1}^{n} \rho_{\alpha }\displaystyle\varepsilon_\alpha$ is the total intrinsic internal energy, $\mathbf{q}_{I}=\sum_{\alpha =1}^{n}\mathbf{q}_{\alpha }$ is the
total intrinsic heat flux and $\bm{\sigma }_{I}=\sum_{\alpha =1}^{n}\mathbf{
\sigma }_{\alpha }\mathbf{\ }$\ is the total intrinsic shear stress.\newline
We obtain by summation of Eqs. (\ref{RT_model}),
\begin{equation}
\left\{
\begin{array}{ll}
& \displaystyle\frac{\partial \rho }{\partial t}+\mathrm{div}\,(\rho \mathbf{%
v})=0, \\
&  \\
& \displaystyle\frac{\partial (\rho \mathbf{v})}{\partial t}+\mathrm{div}%
\,(\rho \mathbf{v}\otimes \mathbf{v}-\mathbf{t})=\mathbf{0}, \\
&  \\
& \displaystyle\frac{\partial \left( \frac{1}{2}\rho v^{2}+\rho \varepsilon
\right) }{\partial t}+\mathrm{div}\,\left\{ \left( \frac{1}{2}\rho
v^{2}+\rho \varepsilon \right) \mathbf{v}-\mathbf{t}\mathbf{v}+\mathbf{q}%
\right\} =0,
\end{array}
\right.  \label{RT_claws}
\end{equation}%
which are the conservation laws of mass, momentum and energy of the
mixture. They are in the same form as for a single fluid.
\newline
 In order to compare the balance equations of  mixtures and  single fluids, we  write   Eqs. (\ref{RT_model}) in the
equivalent form
\begin{equation}
\left\{
\begin{array}{ll}
& \displaystyle\frac{\partial \rho }{\partial t}+\mathrm{div}\,(\rho \mathbf{%
v})=0, \\
&  \\
& \displaystyle\frac{\partial (\rho \mathbf{v})}{\partial t}+\mathrm{div}%
\,(\rho \mathbf{v}\otimes \mathbf{v}-\mathbf{t})=\mathbf{0}, \\
&  \\
& \displaystyle\frac{\partial \left( \frac{1}{2}\rho v^{2}+\rho \varepsilon
\right) }{\partial t}+\mathrm{div}\,\left\{ \left( \frac{1}{2}\rho
v^{2}+\rho \varepsilon \right) \mathbf{v}-\mathbf{t}\mathbf{v}+\mathbf{q}%
\right\} =0, \\
&  \\
& \displaystyle\frac{\partial \rho _{b}}{\partial t}+\mathrm{div}\,(\rho _{b}%
\mathbf{v}_{b})=0,\qquad \qquad (b=1,\dots ,n-1)  \\
&  \\
& \displaystyle\frac{\partial (\rho _{b}\mathbf{v}_{b})}{\partial t}+\mathrm{%
div}\,(\rho _{b}\mathbf{v}_{b}\otimes \mathbf{v}_{b}-\mathbf{t}_{b})=\mathbf{%
m}_{b}, \\
&  \\
& \displaystyle\frac{\partial \left( \frac{1}{2}\rho _{b}v_{b}^{2}+\rho
_{b}\varepsilon _{b}\right) }{\partial t}+\mathrm{div}\,\left\{ \left( \frac{%
1}{2}\rho _{b}v_{b}^{2}+\rho _{b}\varepsilon _{b}\right) \mathbf{v}_{b}-%
\mathbf{t}_{b}\mathbf{v}_{b}+\mathbf{q}_{b}\right\} =e_{b} ,
\end{array}%
\right.  \label{minca}
\end{equation}%
where the index $b$ runs from $0$ to $n-1$.\newline

In this multi-temperature model (\emph{MT}),  used in particular in plasma physics \cite{Bose}, we have $5n$ independent field variables
$\rho _{\alpha },\ \mathbf{v}_{\alpha }$ and $T_{\alpha } \ (\alpha
=1,2,\dots ,n)$, where $T_{\alpha }$ is the temperature of constituent $\alpha$. To close the system (\ref{minca}) of the field equations of the mixture process, we must write the constitutive equations for the quantities $p_{\alpha }, \varepsilon _{\alpha }, \mathbf{q}_{\alpha }, \mathbf{\sigma }_{\alpha }\ (\alpha =1,2,\dots ,n)\ $and $
\mathbf{m}_{b},\ e_{b}\ (b=1,\dots ,n-1)$  in terms of the field variables $\rho_{\alpha }, \mathbf{v}_{\alpha }$ and $T_{\alpha } \ (\alpha
=1,2,\dots ,n)$.

\section{Coarser theories}

Due to the difficulties in measuring the  temperature of each component, a common
practice among engineers and physicists is to consider only one temperature for
the mixture. When we use a single temperature (\emph{ST}), Eq. (\ref{minca})$_6$ disappears and we get a unique global conservation of the total energy in the form (\ref{minca})$_3$ (see for example \cite{ret}).
 In a recent paper, Ruggeri and
Simi\'{c} \cite{SR} discussed the mathematical difference between the \emph{ST} and the \emph{MT} models when the fluid components are
Eulerian gases ($\mathbf{q}_{\alpha }=0, \mathbf{\sigma }_{\alpha }=0$){.
They proved that the differential system of the \emph{ST} model is a \emph{%
principal sub-system} \cite{subsys} of the \emph{MT} model, and for large
times, \emph{MT} solutions converge to \emph{ST} ones.

A further step of coarsening theory is
 the classical approach of mixtures, in which the independent field variables are the density, the mixture velocity, the individual temperature of the mixture and the concentrations of  constituents. \newline
 In that case  system (\ref{minca}) reduces to the equations
\begin{equation}
\left\{
\begin{array}{l}
\dfrac{d\rho }{dt}+\rho \,\text{div}\,\mathbf{v}=0, \\
\\
\rho \dfrac{d\mathbf{v}}{dt}-\func{div}\mathbf{t}=\textbf{0}, \\
\\
\rho \dfrac{d\varepsilon }{dt}-\mathbf{t}\ \func{grad}\,\mathbf{v}+\text{div}%
\,\mathbf{q}=0, \\
\\
\rho \dfrac{dc_{b}}{dt}+\text{div}\,\mathbf{J}_{b}=0,\ \ \ \ (b=1,\cdots,
n-1),
\end{array}%
\right.  \label{material}
\end{equation}
where
\begin{equation*}
\frac{d}{dt}=\frac{\partial }{\partial t}+\mathbf{v\cdot }\frac{\partial }{%
\partial \mathbf{x}}
\end{equation*}
represents the material derivative of the mixture motion,
\begin{equation}
c_{\alpha }=\frac{\rho _{\alpha }}{\rho },  \qquad \left( \sum_{\alpha
=1}^{n}c_{\alpha }=1\right)  \label{concentrations}
\end{equation}
are the components' concentrations, and
\begin{equation}
\mathbf{J}_{\alpha }=\rho
_{\alpha }\mathbf{u}_{\alpha }=\rho _{\alpha }\left( \mathbf{v}_{\alpha }-%
\mathbf{v}\right), \qquad \left( \sum_{\alpha =1}^{n}\mathbf{J}_{\alpha
}=0\right)    \label{diffJ}
\end{equation}
are the diffusion fluxes of the components.
\newline
In the classical approach the stress tensor - as  in a single
fluid - splits up into the pressure isotropic part and the viscosity
stress tensor $\mathbf{ \sigma}$ (for Stokesian fluids this is a
deviatoric tensor)
\[
\mathbf{t} = -p \mathbf{I} +\mathbf{ \sigma}.
\]
The system (\ref{material}) determines the field variables $\rho
,  T, \mathbf{v} $ and $c_{b}\ (b=1,\cdots, n-1)$. Consequently, we need constitutive relations for $\varepsilon ,\mathbf{ \sigma}, \mathbf{q}$ and $\mathbf{J}_{b}  \ (b=1,\cdots, n-1)$. \newline
We consider the pressure $p(\rho ,T,c_{b})$ and the internal
energy $\varepsilon (\rho ,T,c_{b})$ as given by the equilibrium equations
of state as they appear in the Gibbs equations for mixture, viz.
\begin{equation}
TdS=d\varepsilon -\frac{p}{\rho ^{2}}\,d\rho -\sum_{b=1}^{n-1}\left( \mu
_{b}-\mu _{n}\right) \ dc_{b}  \label{gibbseq}
\end{equation}%
where $\mu _{\alpha }=\mu _{\alpha }(\rho ,T,c_{b})$,  with $ \alpha=1,\cdots, n $, denote the chemical
potentials of the components at equilibrium and $S$ is the entropy density of the mixture. \newline
The entropy balance law is a consequence of equation (\ref{gibbseq}) and system
(\ref{material}). By using arguments from the thermodynamics of irreversible processes \emph{(TIP)} presented in  \cite{Eckart} and \cite{ret}
chapter 5, we obtain  the classical constitutive equations of mixtures
\begin{eqnarray}
&&\mathbf {\sigma}=2\,\nu\, \textbf{D}^D , \notag \\
&&\mathbf {q}=L \ \text{grad} \left(\frac{1}{{T}}\right)
+\sum_{b=1}^{{n} -1} {L}_{{b}}\ \text{grad} \left(  \frac{\mu
_{{b}}-\mu _{{n}}} {{{T}}}\right),  \label{1.9} \\
&&\mathbf {J}_{{a}}=\tilde{L}_{{a}}\ \text{grad}\left(\frac{1}{{T}}\right) -\sum_{{b} =1}^{{n} -1}{L}_{{a}{b} } \ \text{grad}   \left(  \frac{\mu
_{{b}}-\mu _{{n}}} {{{T}}}\right)   \notag
\end{eqnarray}%
where $\textbf{D}^D$   denotes    the deviatoric part of the strain
velocity tensor $\textbf{D} = \frac{1}{2}\left(\nabla \textbf{v}
+(\nabla \textbf{v})^T\right)$. The \emph{phenomenological
coefficients} $L, L_{b},\tilde{L}_a$ and $L_{a b}\, (a, b=1,\cdots
,n-1)$ are the transport coefficients of heat conduction and
diffusion.\\ Let us note that relation (\ref{1.9})$_{1}$  is the
classical  Navier-Stokes equation  of a Newtonian (Stokesian and
isotropic) fluid, while (\ref{1.9})$_{2,3}$ are generalizations of
the original phenomenological laws of Fourier and Fick, according to
which the heat flux and the diffusion flux depend on the gradients
of temperature and concentrations respectively (\emph{but not on
both}). The \emph{TIP} permits the temperature gradient to influence the
diffusion fluxes and concentration gradients to influence the heat
flux; both effects are indeed observed and they are called,
respectively, thermo-diffusion  and diffusion-thermo  or
Soret effects. Additionally, the Onsager conditions of symmetry yield
the following symmetries of coefficients \cite{onsag}
\begin{equation}
\begin{array}{ll}
L_{ab}=L_{ba}\, , \  & \ \ \tilde{L}_{b}= L_{b}\ \ (a, b=1,\cdots , n-1)
\end{array} \label{positive}
\end{equation}
and the following inequalities must be satisfied:
\begin{eqnarray}
&&\left[
\begin{array}{cc}
L & L_{b} \\
\tilde{L}_a & L_{a b}%
\end{array}%
\right]\qquad \hbox{is a positive definite form,} \notag \\
\label{coeff}
\\
&& \         \text{and} \quad \nu \geq 0, \notag
\end{eqnarray}
so that the entropy inequality can be satisfied.

\section{A Classical approach to multi-temperature mixtures and  the  average temperature}

To reveal the relation between the extended and classical models, a
formal iterative scheme known as \emph{Maxwellian iteration}  is
applied (see e.g \cite{ret}). In the case of the $ST$ model
the first iterates $\mathbf{J}_{a}^{(1)}$  are  calculated from the right--hand sides of the balance laws (\ref{minca})$_5$ by using "zeroth" iterates -
equilibrium values $\mathbf{J}_{a}^{(0)} = \mathbf{0}$ - on the left-hand sides. The next step, second iterates
$\mathbf{J}_{a}^{(2)}$ are obtained from the
right--hand sides of the same equations  by putting first iterates
$\mathbf{J}_{a}^{(1)}$  on their left--hand sides, an
so on. If we apply the  first \emph{Maxwellian iteration}
 the Fick laws
 of diffusion fluxes (\ref{1.9})$_3$ are obtained. Roughly speaking, the Fick laws are obtained by neglecting the accelerations of the relative
motions of the constituents and the classical theory is an approximation of the $ST$ model (see \cite{ret}, Chapter $5$).\newline
In a recent paper, Ruggeri and Simi\'{c}   \cite{Greco}  considered the Maxwellian iteration
 of  system (\ref{minca}) in the case of a binary mixture of Eulerian fluids; they
obtained the Fick laws  as a first order term of the expansion of the component momentum equations. When each component has its own temperature, an additive constitutive equation comes from a limiting case of the constituent equation of energy in the form
\begin{equation}
\Theta =L_{\theta }\,(\gamma _{1}-\gamma _{2})\,\mathrm{div}\,\mathbf{v} \label{tetaRS}
\end{equation}%
 where $L_{\theta }$ is a new phenomenological positive coefficient and
\begin{equation*}
 \Theta =T_{2}-T_{1}\,.
 \end{equation*}
Eq. (\ref{tetaRS}) is not obtained  in   classical theory. The aim
of this paper is to find a variant form of classical approach to
recover also equations like  (\ref{tetaRS}) in the general case of
mixtures with $n$ not necessarily Eulerian compressible
fluids.\newline

In the classical approach, the velocity field $\mathbf{v}$
corresponds to an average   velocity with respect to mass components. Thanks to the Fick Laws,
 the diffusion fluxes ${\bf J}_\alpha=\rho_\alpha\left({\bf v}_\alpha -{\bf v}\right)$ determine the component velocities
  ${\bf v}_\alpha$.

By analogy with  the velocity fields  a natural extension of the
classical approach is to consider   an average temperature $T$ and
$\Theta_\alpha=T_\alpha -T  \ (\alpha =1, \cdots, n)$
 as constitutive quantities; similarly with the diffusion velocity fluxes, we name $\Theta_\alpha$
 \emph{the diffusion temperature fluxes}.\newline
To define an average temperature $T$, Ruggeri and Simi\'{c} assumed that
the total intrinsic internal energy of the mixture (which coincides with the full internal energy for processes not so far from equilibrium (see (\ref{6})))  is the same in the
multi-temperature case as in the
 \emph{ST} model when the temperature is $T$ \cite{Renno}.
Consequently, $T$ is defined through the local implicit solution of
the equation
\begin{equation}
\rho \varepsilon \equiv \sum_{\alpha =1}^{n}\rho _{\alpha }\varepsilon
_{\alpha }(\rho _{\alpha },T)=\sum_{\alpha =1}^{n}\rho _{\alpha }\varepsilon
_{\alpha }(\rho _{\alpha },T_{\alpha }) .  \label{temperature}
\end{equation}
This choice comes from the case of particular classes of solutions
for perfect gases \cite{Renno} and from  the fact   that the
equation of energy governs the evolution of the common  temperature $T$ for the \emph{ST}
model. The consequences on the entropy of the mixture will confirm the physical grounds   of Eq. (\ref{temperature}).\newline
Taking into account Eqs. (\ref
{material})$_{1}$, $(\ref{material})_{4}$  and  (\ref{temperature}),      Eq. (\ref{material})$_{3}$ of energy of the mixture  can be written  as  a
differential equation for  the average temperature $T$
\begin{equation}
\rho \varepsilon _{,T}\frac{dT}{dt}=\rho ^{2}\varepsilon _{,\rho }\,\text{div}
\,\mathbf{v+}\sum_{b=1}^{n-1}\varepsilon _{,c_{b}}\mathrm{div}\,\mathbf{J}%
_{b} + \mathbf{t} \  \text{grad}\, \mathbf{v} - \text{div\ }\mathbf{q} , \label{dT}
\end{equation}
where the comma denotes the partial derivative with respect to the subscript.
\newline
As
 $\mathbf{J}_\alpha\ (\alpha = 1, \cdots, n)$ are associated with the difference between component and average velocities
\begin{equation*}
\Theta _{\alpha }=T_{\alpha }-T\quad (\alpha = 1, \cdots, n)
\end{equation*}
corresponding to the difference between component and average temperatures, are non-equilibrium thermodynamical variables.
Near the equilibrium,   Eq. (\ref{temperature}) can be expanded to the
first order; then
\begin{equation}
\sum_{\alpha =1}^{n}\rho _{\alpha }c_{V}^{(\alpha )}\Theta _{\alpha }=0
\quad \Longleftrightarrow \quad T=\dfrac{\sum_{\alpha =1}^{n}\rho _{\alpha
}c_{V}^{(\alpha )}T_{\alpha }}{\sum_{\alpha =1}^{n}\rho _{\alpha
}c_{V}^{(\alpha )}}   \label{newT}
\end{equation}
where
\begin{equation}
c_{V}^{(\alpha)}=\frac{\partial \varepsilon _{\alpha}}{\partial T_{\alpha}}%
\left( \rho _{\alpha},T\right)  \label{cvalpha}
\end{equation}
denotes the specific heat at constant volume for the constituent ${\alpha}$ at  equilibrium. Consequently,  Eq. (\ref{newT}) yields
\begin{equation}
\Theta _{n}=-\frac{1}{\rho _{n}c_{V}^{(n)}}\sum_{b=1}^{n-1}\rho
_{b}\,c_{V}^{(b)}\Theta _{b}  . \label{alfetta2}
\end{equation}
The definition of the total specific entropy $S$ of the mixture is
\begin{equation}
\rho S=\sum_{\alpha =1}^{n}\rho _{\alpha }S_{\alpha }(\rho _{\alpha
},T_{\alpha }) ,  \label{entropia}
\end{equation}
where $S_\alpha  \ (\alpha = 1, \cdots, n)$ are the specific entropies of the components. Let us note that the specific entropy $S$ depends only on $T$ and not on $\Theta _{b} \, \, (b=1,\dots,n-1)$.
This
property comes from the Gibbs relation of each constituent,
\begin{equation*}
T_{\alpha }dS_{\alpha }=d\varepsilon _{\alpha }-\frac{p_{\alpha
}}{\rho _{\alpha }^{2}}d\rho _{\alpha }\, ,
\end{equation*}
which implies
\begin{equation}
T_{\alpha }\frac{\partial S_{\alpha }}{\partial T_{\alpha }}\left( \rho
_{\alpha },T_{\alpha }\right) =\frac{\partial \varepsilon _{\alpha }}{
\partial T_{\alpha }}\left( \rho _{\alpha },T_{\alpha }\right)   . \label{Sei}
\end{equation}
The first-order expansion of Eq.(\ref{entropia}) yields
\begin{equation*}
 \rho\ S =\sum_{\alpha =1}^{n}\left\{ \rho _{\alpha }S_{\alpha }(\rho _{\alpha
},T)+\rho _{\alpha }\frac{\partial S_{\alpha }}{\partial T_{\alpha }}\left(
\rho _{\alpha },T\right) \ \Theta _{\alpha }\right\}   .
\end{equation*}
Equation (\ref{Sei}) can be evaluated for $T_{\alpha }=T$; by using Eq. (\ref{newT}) we deduce
\begin{equation*}
\sum_{\alpha =1}^{n}  \rho _{\alpha }\frac{\partial S_{\alpha }}{\partial T_{\alpha }}\left(
\rho _{\alpha },T\right) \ \Theta _{\alpha } =  0
\end{equation*}
and consequently
\begin{equation}
\rho S = \sum_{\alpha =1}^{n}\rho _{\alpha }S_{\alpha }(\rho _{\alpha },T)   .
\label{SSS}
\end{equation}
The specific entropy
$S$ does
not depend on $\Theta_b\ (b = 1, \cdots, n-1) $.\newline
On the contrary, a first-order expansion of the total pressure of the mixture $p=\sum_{\alpha=1}^n p_\alpha(\rho_\alpha,T_\alpha)$ together with
Eq. (\ref{alfetta2})
yields
\begin{equation}
p=p_{0}+\pi_\theta,  \label{pressionet}
\end{equation}
where
\begin{equation} \label{paiteta}
p_{0}=\sum_{\alpha =1}^{n}p_{\alpha }(\rho _{\alpha },T), \quad \pi_\theta = \sum_{b=1}^{n-1}r_{b}\, \Theta _{b}
\end{equation}
and
\begin{equation}
r_{b}=\frac{1}{\rho _{n}c_{V}^{(n)}}\left\{ \rho _{n}c_{V}^{(n)}\frac{%
\partial p_{b}}{\partial T_{b}}\left( \rho _{b},T\right) -\rho
_{b}c_{V}^{(b)}\frac{\partial p_{n}}{\partial T_{n}}\left( \rho
_{n},T\right) \right\}  \quad (b=1,\dots,n-1).   \label{rrr}
\end{equation}
Therefore,  the total pressure $p$ of the mixture is a sum of the
equilibrium part $p_0$  depending on
 $T$ and  a new dynamical pressure part (as a non-equilibrium term) $\pi_\theta$
 due  to the difference of temperatures between the constituents.

We emphasize that the fact that the entropy density depends only on $T$ and not on the $\Theta_a$ justifies - in the present theory -    the consideration of $\Theta_a$  as constitutive quantities (undetermined quantities). This is a key point in the model. \\ Consequently, the aim of the next section consists in determining the constitutive equations for the $\Theta_a$  by  using  the entropy principle and \emph{TIP} arguments.

\section{Entropy principle and  constitutive equations}
We still assume that the internal energy $\varepsilon(\rho,T,c_b)$
 and the
equilibrium pressure $p_{0}(\rho,T,c_b)$ satisfy the Gibbs equation
\begin{equation}
TdS=d\varepsilon -\frac{p_{0}}{\rho ^{2}}\,d\rho
-\sum_{{b}=1}^{{n}-1}\left( \mu _{b}-\mu _{n}\right) \ dc_{b}\, .
\label{newGibbs}
\end{equation}
The  differences between  Eq.  (\ref{gibbseq}) and  Eq.
(\ref{newGibbs}) consist in the fact that in Eq. (\ref{newGibbs})
\emph{T} means the average temperature when  each component $\alpha$
has its own temperature $T_\alpha$   and   $p_0 $ takes the place of
$p$\,. At equilibrium, the component chemical potentials are
\begin{equation*}
\mu _{\alpha }=\varepsilon _{\alpha }(\rho _{\alpha },T)+\frac{p_{\alpha
}(\rho _{\alpha },T)}{\rho _{\alpha }}-TS_{\alpha }(\rho _{\alpha },T).
\end{equation*}
As in the classical case with a single temperature,
 the time derivative can be eliminated between
 Eqs (\ref{material})$_{1}$, (\ref{material})$_{4}$ and (\ref{dT}).
  The second-order terms due to the diffusion velocities $\mathbf{u}_\alpha \ (\alpha= 1, \cdots, n)$
  are neglected; because we  consider  a Stokesian fluid, the viscous stress tensor is deviatoric
   and we obtain the balance law in the form

\begin{eqnarray}
& & \rho \frac{dS}{dt}+\text{div}  \left\{  \frac{1}{T}\left( \mathbf{q} -\sum_{{b}=1}^{{n }-1}(\mu _{b}-\mu
_{n})\,\mathbf{J}_{b}\right)\right \} = \notag \\
&& \mathbf{q}\cdot \text{grad} \left(\frac{1}{{T}}\right)
-\sum_{{b}=1}^{{n}-1}\mathbf{J}_{{b}}\cdot \text{grad}   \left(
\frac{\mu _{{b}}-\mu _{{n}}} {{{T}}}\right)  +
\frac{1}{{T}}\,\text{tr}\left(\mathbf{J}_{mech}  \, {\mathbf D}
\right),  \label{entro1}
\end{eqnarray}
where the mechanical flux (for Stokesian fluids) is
\begin{equation}
\mathbf{J}_{mech}  = \mathbf{\sigma}-
\pi_\theta\,\textbf{I}.
\end{equation}
Eq. (\ref{entro1}) can be interpreted as a balance of entropy,
if we
consider\begin{equation*} {\bm \Phi} = \frac{1}{T}\left( \mathbf{q}
-\sum_{{b}=1}^{{n }-1}(\mu _{b}-\mu _{n})\,\mathbf{J}_{b}\right)
\end{equation*}
and
\begin{eqnarray}
&& \Sigma = \mathbf{q}\cdot \text{grad} \left(\frac{1}{{T}}\right)
-\sum_{{b}=1}^{{n}-1}\mathbf{J}_{{b}}\cdot \text{grad}   \left(
\frac{\mu _{{b}}-\mu _{{n}}} {{{T}}}\right)  +
\frac{1}{{T}}\,\text{tr}\left(\mathbf{J}_{mech}  \, {\mathbf D}
\right) \label{25}
\end{eqnarray}
as the entropy flux and the entropy production respectively.

We
observe that the entropy production is the sum of products of the
following quantities: {\normalsize
\begin{equation*}
\begin{array}{lll}
\mbox{\bf thermodynamic fluxes} &   & \mbox{\bf thermodynamic forces} \\

\mbox{heat flux}\,\,  \mathbf{q} &  & \mbox{temperature gradient}\,\, \text{grad}\left(
 \frac{1}{{T}}\right),  \\

 \mbox{diffusion fluxes}\,\,  \mathbf{J}_b &  &
\mbox{chemical potential
gradients}\,\, \text{grad}   \left(  \frac{\mu
_{b}-\mu _{{n}}} {{{T}}}\right), \\

\mbox{mechanical flux}\,\, \mathbf{J}_{mech}     & & \mbox{velocity
gradient}\,\,\textbf{D}.
\\
%
\end{array}%
\end{equation*}

 In accordance with the case of a single temperature
model \cite{mullermixture,mullerthermody} and \cite{ret} chapter 5,
in \emph{TIP} near equilibrium, the fluxes depend linearly on the
associated forces (see also for the general methodology of the
\emph{TIP} \cite{Eckart,onsag,de Groot,Gyarmati}):

- For the heat flux and  the diffusion fluxes, we obtain the
constitutive equations in the form of Eqs (\ref{1.9})$_{2,3}$.

- For Stokesian fluids, the last term of Eq. (\ref{25})  corresponding to
the mechanical production of entropy can be written
 in a    separated form as
\begin{equation*}
 \frac{1}{{T}}\,\text{tr}\left(\mathbf{J}_{mech}  \, {\mathbf D}
\right) = \frac{1}{{T}}\,\text{tr}\left(\mathbf{\sigma}  \, {\mathbf
D}^{D} \right) {- \frac{1}{T}\,  \pi_\theta \,
\text{div} \,   \mathbf{v}} . \label{entro}
\end{equation*}
We obtain the constitutive equation of the viscous stress tensor  in the form of
 Eq.
(\ref{1.9})$_{1}$ and
  the dynamical pressure part due to the difference of temperatures
  yields
\begin{equation}\label{ppteta}
\pi_\theta = \sum_{b=1}^{n-1}r_{b}\, \Theta _{b}=-L_{\pi }\,
\text{div}\,\mathbf{v},
\end{equation}
where $L_\pi$ is a scalar coefficient of proportionality.

The production of entropy must be  non-negative \cite{CN,IM} and therefore   the \emph{phenomenological coefficients} must
satisfy the inequalities (\ref{coeff}) and
\begin{equation}
L_\pi \geq 0.
\end{equation}
Taking into account that the terms $r_b$ given by Eq. (\ref{rrr}) depend  on $(\rho_b,T)$,
 from Eq. (\ref{ppteta})  we deduce  that the  constitutive quantities
 $\Theta_a$ (depending \emph{a priori}  on $\nabla \mathbf{v}$)  must be proportional
 to $\text{div}\,\mathbf{v}$:
\[
\Theta_a = k_a \, \text{div}\,\mathbf{v} \quad (a =1, \cdots, n-1).
\]
Let $\left\|M_{ab}\right\|$ be the matrix
such that $k_a = \sum_{b=1}^{n-1}M_{a b}\ r_{b}$, we have
\begin{equation}
\Theta _a  =-\sum_{b=1}^{n-1}M_{a b}\ r_{b}\,\text{div}\,\mathbf{v}\quad (a =1, \cdots, n-1).
\label{new}
\end{equation}
Introducing expression (\ref{new}) into Eq. (\ref{ppteta}), we obtain
\[
L_{\pi }=\sum_{a,b=1}^{n-1}M_{ab}\ r_{a}r_{b}\geq 0 ,
\]
and  assuming
the  Onsager
 symmetry property,
$ M_{a b}= M_{b a}\  (a, b = 1, \cdots, n-1) $, we deduce that the  coefficients $ M_{a b}$ are
\emph{associated with a positive definite quadratic form}.
\newline
Finally, the results are
 the same as in   classical theory, but in addition we get     new
constitutive equations (\ref{new}) for the difference of  temperatures.

We have considered the simple case of Stokes fluids. If the fluid is non-Stokesian,   the Navier-Stokes stress tensor of viscosity is
\begin{equation*}
\mathbf{\sigma }=\lambda \ (\text{div}\,\mathbf{v})\,\mathbf{I}+2\,\nu \
\mathbf{D}^{D},
\end{equation*}%
where $\lambda$ is the bulk viscosity.   The stress
tensor $\mathbf{t}$   becomes
\begin{equation*}
\mathbf{t}=-(p_{0}+\pi _{\theta })\,\mathbf{I}+\mathbf{\sigma }=-p\
\mathbf{I}+2\,\nu \ \mathbf{D}^{D},
\end{equation*}%
with
\begin{equation*}
p=p_{0}+\pi _{\theta }+\pi _{\sigma }      .
\end{equation*}%
The nonequilibrium pressure $p-p_{0}$ is separated into two
different parts. The first one $\pi _{\sigma }=-\lambda \ \text{div}\,%
\mathbf{v}\ $ is related to the bulk viscosity and the second one $\pi
_{\theta }=-L_{\pi }\,\text{div}\,\mathbf{v}\ $ is related to the multitemperature
effects between components.

For example, we consider the case of mixtures of perfect gases
\begin{equation*}
p_{\alpha }=\frac{k}{m_{\alpha }}\ \rho _{\alpha }T_{\alpha },\ \ \ \ \ \
\varepsilon _{\alpha }=c_{V}^{(\alpha )}\,T_{\alpha }\quad \text{with}\quad
c_{V}^{(\alpha )}=    \frac{k}{{m}_{\alpha }\left( \gamma _{\alpha
}-1\right) },
\end{equation*}%
where $k$ and $m_{\alpha }$ are respectively the Boltzmann constant and the
atomic mass of constituent $\alpha $. From Eq. (\ref{rrr}), we get
\begin{equation*}
r_{b}=\rho _{b}\,c_{V}^{(b)}\left( \gamma _{b}-\gamma _{n}\right)
\end{equation*}%
and consequently
\begin{equation}
\Theta _{a}\ =-\sum_{b=1}^{n-1}M_{ab}\ \rho _{b}\,c_{V}^{(b)}\left( \gamma
_{b}-\gamma _{n}\right) \,\mathrm{div}\,\mathbf{v}  . \label{tetissima}
\end{equation}%
The dynamical pressure is
\begin{equation}
\pi _{\theta }=-L_{\pi }\,\text{div}\,\mathbf{v}\quad \text{with \ }L_{\pi
}=\sum_{a,b=1}^{n-1}M_{ab}\ \rho _{a}\rho _{b}\,c_{V}^{(a)}c_{V}^{(b)}\left(
\gamma _{a}-\gamma _{n}\right) \left( \gamma _{b}-\gamma _{n}\right)         .
\label{dasi}
\end{equation}
Let us note that Eq. (\ref{tetissima}) yields the result of Ruggeri and Simi\'{c}
\cite{Greco} for binary mixtures:
\begin{equation}
\Theta =L_{\theta }\,(\gamma _{1}-\gamma _{2})\,\mathrm{div}\,\mathbf{v}%
\quad \text{with}\quad L_{\theta }=\ M_{11}\ \frac{\rho _{1}\,c_{V}^{(1)}}{%
\rho _{2}\,c_{V}^{(2)}}\left( \rho _{1}c_{V}^{(1)}+\rho
_{2}c_{V}^{(2)}\right)  . \label{moi}
\end{equation}%
\section{Kinetic values of the phenomenological coefficients}
While the \emph{TIP} yields only inequalities for the phenomenological  coefficients, they can be determined by the fact that our approach is a limiting case of   system (\ref{minca}) which is in agreement with kinetic theory.
\newline
For high temperature plasma physics \cite{Bose}, the phenomenological
coefficients appear through the production terms $\mathbf{m}_{b}$ and $e_{b}$ of
Eqs.(\ref{minca}) and have explicit expressions. Therefore by using the Maxwellian iteration, we can evaluate
coefficients $L_{ab}$ and $M_{ab}$. For a binary mixture of Eulerian gases the calculus is
done in \cite{Greco} and \cite{Renno} and we get
\begin{equation}
L_{11}=\tau _{J} \ \frac{\rho _{1}\rho _{2}}{\rho }  \, T \quad  {\text{and}} \quad L_{\theta
}=\tau _{\theta }\, T   \label{elles}
\end{equation}%
where $\tau _{J}$ and $\tau _{\theta }$ are, respectively, the relaxation times of the
\emph{mechanical diffusion} and the \emph{multi-temperature diffusion}
\begin{eqnarray}
\tau _{J} &=&\frac{1}{2\Gamma _{12}^{^{\prime }}}\frac{m_{1}+m_{2}}{%
m_{1}m_{2}}\ \ \frac{\rho _{1}\rho _{2}}{\rho }\ ,    \notag \\
\ \tau _{\theta } &=&\frac{1}{3\Gamma _{12}^{^{\prime }}}\frac{%
(m_{1}+m_{2})^{2}}{m_{1}m_{2}}\frac{\rho _{1}\rho _{2}}{\rho
_{1}m_{2}(\gamma _{2}-1)+\rho _{2}m_{1}(\gamma _{1}-1)}\ ,    \label{tempis}
\end{eqnarray}%
where $\Gamma _{12}^{^{\prime }}$ is the number of collisions per units of
volume and time. The coefficient $L_{\pi }$ of the dynamical pressure $\pi _{\theta }
$ in Eq. (\ref{dasi}) is directly obtained from Eqs. (\ref{moi})$_{2}$, (\ref%
{elles}) and (\ref{tempis})
\begin{equation*}
L_{\pi }=\frac{kT\rho _{1}^{2}\,\rho _{2}^{2}}{3\,\Gamma _{12}^{^{\prime }}\
m_{1}m_{2}}\ \frac{(m_{1}+m_{2})^{2}}{\left[ \rho _{1}m_{2}(\gamma
_{2}-1)+\rho _{2}m_{1}(\gamma _{1}-1)\right] ^{2}}\;\left( \gamma
_{1}-\gamma _{2}\right) ^{2}             .
\end{equation*}%
This coefficient is null only if the two gases have the same $\gamma _{1}$
and $\gamma _{2}$ values.\newline
The ratio of the two relaxation times is
\begin{equation*}
\frac{\ \tau _{\theta }}{\tau _{J}}=\frac{2}{3}\;\frac{(m_{1}+m_{2})\rho }{%
\rho _{1}m_{2}(\gamma _{2}-1)+\rho _{2}m_{1}(\gamma _{1}-1)}          .
\end{equation*}%
This ratio is always greater than $1$. In fact  if we assume
that ${\gamma _{1}\neq \gamma _{2}}$ and denote by $\gamma _{\max }=\max {
(\gamma _{1},\gamma _{2})}$, we get ($c = \rho_1/\rho$)
\begin{equation*}
\frac{\ \tau _{\theta }}{\tau _{J}}>\frac{2}{3}\,\frac{(m_{1}+m_{2})}{%
cm_{2}+(1-c)m_{1}}\frac{1}{(\gamma _{\max }-1)}>\frac{2}{3}\,\frac{1}{%
(\gamma _{\max }-1)}\geq 1
\end{equation*}%
(the last inequality is because ${\gamma _{\alpha }}$ ${\leq 5/3}$ for all
ideal gases).
\newline
In contrast with the usual approach, in which we consider
diffusion and   neglect the difference of temperatures between constituents, the
relaxation time of the multitemperature diffusion is always greater than the
relaxation time of classical diffusion. \newline
From Eq. (\ref{elles}), we can see that
if $\tau _{\theta }$ is small enough, to obtain $L_{\theta }$ and $L_{11}$ values
close to unity, it is necessary to have a high temperature $T$. Consequently,  in  the high-temperature cases,   the multitemperature diffusion terms cannot be neglected.

\section{Conclusion}

In the case of fluid mixtures with different temperatures, a
classical framework allows one to determine  the novel constitutive equation (\ref{new}) for the
difference of temperatures in addition to  the classical
Navier-Stokes, Fourier and Fick laws. The difference of the temperatures produces a new dynamical  pressure
term which has a physical meaning and consequently can be measured. We
point out that the dynamical pressure associated with
multitemperature fluid mixtures exists even if the bulk
viscosity is null. For rarefied gases, it is well known that the bulk viscosity $%
\lambda $ is null (Stokes fluids) and consequently no dynamical pressure
exists in a classical mixture with one temperature; in a relativistic
context, this pressure exists but remains very small \cite{ret}.
Due to the
nonzero dynamical pressure even for Stokes fluids, we
 conclude
that multitemperature mixtures of fluids have a great importance.
\newline
Perhaps such a model may be
used to analyze the evolution of the early universe in which a dynamical
pressure seems essential \cite{Groot,Weinberg}. Of course, we have to take into account a
relativistic framework and the chemical reactions. Nevertheless, the preliminary result - \emph{a new
dynamical pressure exists also in the non-relativistic limit and without
chemical reactions} - shows  promise and  deserves attention for
 future work.
\newline
We focus attention on the fact that the present results are obtained
thanks to an average temperature corresponding to the same internal
energy as for the single-temperature model. \newline It is
important to note that the total energy conservation law yields the
evolution of $T$, the entropy principle can be used because
the entropy density $S$ depends only on $T$ and the Gibbs equation
is always defined by using Eq. (\ref{newGibbs}). \newline Finally if
we consider Eq. (\ref{dT}) for the field variables to be
spatially homogeneous, $T$ is constant as we expect from physical
point of view (see ref. \cite{Renno}).

\bigskip{\small {\ \textbf{Acknowledgment}: {
The authors thank the     anonymous referees for the  criticisms
allowing   improvement of the paper and are very grateful to
Ingo M\"uller for stimulating discussions.
\newline
This paper was developed during
a stay of T.R. as visiting professor in U.M.R. 6181 at the University of
Aix-Marseille and was supported in part (T.R.) by fondi MIUR Progetto di
interesse Nazionale \emph{Problemi Matematici Non Lineari di Propagazione e
Stabilit\`{a} nei Modelli del Continuo}, Coordinatore T. Ruggeri, of the
GNFM-INdAM. }}}

\end{document}